\documentclass[preprint,showpacs,amsmath,amssymb]{revtex4}
\usepackage{graphicx}

\begin{document}

\title{The CSM extension for description of the positive and negative parity bands in even-odd nuclei}

\author{A. A. Raduta$^{a),b),c)}$,  C. M. Raduta$^{b)}$ and Amand Faessler$^{d)}$}

\address{$^{a)}$ Department of Theoretical Physics and Mathematics, Bucharest University, Bucharest, POBox MG11, Romania}

\address{$^{b)}$ Department of Theoretical Physics, Institute of Physics and Nuclear Engineering, Bucharest, POBox MG6, Romania}

\address{$^{c)}$ Academy of Romanian Scientists, 54 Splaiul Independentei, Bucharest 050094, Romania}

\address{$^{d)}$Institut f\"{u}r Theoretische Physik der Universit\"{a}t T\"{u}bingen, Auf der Morgenstelle 14, Germany}

\begin{abstract}A particle-core Hamiltonian is used to describe the lowest parity partner bands $K^{\pi}=1/2^{\pm}$ in
 $^{219}$Ra, $^{237}$U and $^{239}$Pu, and three parity partner bands, $K^{\pi}=1/2^{\pm}, 3/2^{\pm}, 5/2^{\pm}$, in $^{227}$Ra.
The core is described by a quadrupole and octupole boson  Hamiltonian which was  previously used for the description of four positive and four negative parity bands in the neighboring even-even isotopes.  The particle-core Hamiltonian consists of four terms: a quadrupole-quadrupole, an octupole-octupole, a spin-spin and a rotational $\hat{I}^2$ interaction, with $\hat {I}$ denoting the total angular momentum.
The single particle space for the odd nucleon consists of three spherical shell model states, two  of positive and one of negative parity. The product of these states with a collective deformed ground state and the intrinsic gamma band state generate, through angular momentum projection, the bands with $K^{\pi}=1/2^{\pm},3/2^{\pm},5/2^{\pm}$, respectively. In the space of projected states one calculates the energies of the considered bands.  The resulting  excitation energies are compared with the corresponding experimental data as well as with those obtained with other approaches. Also, we searched for some signatures for a static octupole deformation in the considered odd isotopes. The calculated branching ratios in $^{219}$Ra agree quite well with the corresponding experimental data.  
\end{abstract}


\maketitle
\renewcommand{\theequation}{1.\arabic{equation}}
\setcounter{equation}{0}

\section{Introduction}
The coherent state model (CSM)\cite{Rad81} describes in a realistic fashion three interacting bands, ground, beta and gamma, in terms of quadrupole bosons. The formalism was later extended \cite{Rad97,Rad02,Rad03,Rad003,Rad06,Rad006}
 by considering the octupole degrees of freedom. The most recent extension describes eight rotational bands, four of positive  and four of negative parity. Observable like excitation energies, intraband E2 and interband E1, E2 and E3 reduced transition probabilities have been calculated and the results were compared with the corresponding experimental data. The formalism works well for both
near spherical and deformed nuclei excited in low and high angular momentum states. Indeed, we considered all states with $J\le 30$ in 
both, the positive and the negative parity bands. Signatures for a static octupole deformation in ground as well as in excited bands have been pointed out in several even-even nuclei.

The aim of this paper is to extend CSM for the even-odd nuclei which exhibit both quadrupole and octupole deformation.

The formalism concerning the excitation energies in the positive and negative parity bands is presented in Sections II and III. The  E1 and E2 transitions are considered in Section IV,  while the numerical application to four even-odd nuclei, is described in Section V. The final conclusions are drawn in Section VI.

\renewcommand{\theequation}{2.\arabic{equation}}
\setcounter{equation}{0}

\section{The model Hamiltonian}
We suppose that the rotational bands in even-odd nuclei may be described by a particle-core Hamiltonian:
\begin{equation}
H=H_{sp}+H_{core}+H_{pc},
\label{Hamodd}
\end{equation}
where  $H_{sp}$ is a spherical shell model Hamiltonian associated to the odd nucleon, while $H_{core}$ is a phenomenological Hamiltonian which describes the collective motion of the core in terms of quadrupole and octupole bosons. This term is identical to that used in Ref.\cite{Rad006} to describe eight rotational bands in even-even nuclei. The two subsystems interact with each other by $H_{pc}$, which has the following expression:
\begin{eqnarray}
H_{pc}=&-&X_2\sum_{\mu}r^2Y_{2,-\mu}(-)^{\mu}\left(b^{\dagger}_{2\mu}+(-)^{\mu}b_{2,-\mu}\right)\nonumber\\
       &-&X_3\sum_{\mu}r^3Y_{3,-\mu}(-)^{\mu}\left(b^{\dagger}_{3\mu}+(-)^{\mu}b_{3,-\mu}\right)\nonumber\\
       &+&X_{jJ}\vec{j}\cdot\vec{J}+X_{I^2} \vec{I}^2.
\end{eqnarray}
 $b^{\dagger}_{\lambda \mu}$ denotes the components of the  $\lambda$-pole (with $\lambda$=2,3) boson operator.
The term $\vec{j}\cdot\vec{J}$ is similar to the spin-orbit interaction from the shell model and expresses the interaction between the angular momenta of the odd-particle and the core. The last term is due to the rotational motion of the whole system, $\vec{I}$ denoting the total angular momentum of the particle-core system.

The core states are described by eight sets of mutually orthogonal functions, obtained by projecting out the angular momentum and the parity from four quadrupole and octupole deformed functions: one is a product of two coherent states:
\begin{equation}
\Psi _{g}=e^{ f(b_{30}^{+}-b_{30})}e^{d(b^+_{20}-b_{20})}|0\rangle _{2}|0 \rangle _{3} \equiv \Psi_{o}\Psi_{q}
|0\rangle _{2} | 0\rangle _{3},
\label{cohq2q3}
\end{equation}
while the remaining three are polynomial boson excitations of $\Psi_{g}$. The parameters $d$ and $f$ are real numbers and simulate the quadrupole and octupole deformations, respectively. The vacuum state for the $\lambda$-pole boson, $\lambda=2,3$, is denoted by $|0\rangle_{\lambda}$.

The particle-core interaction generates a deformation for the single particle trajectories. Indeed, averaging the model Hamiltonian with
 $\Psi_{g}$, one obtains a deformed single particle Hamiltonian, $H_{mf}$ which plays the role of the mean field for the particle motion:
\begin{equation}
H_{mf}={\cal C}+H_{sp}-2dX_2 r^2Y_{20}-2fX_3 r^3Y_{30},
\end{equation}
where ${\cal C}$ is a constant determined by the average of $H_{core}$. The Hamiltonian $H_{mf}$ represents an extension of the  Nilsson Hamiltonian by adding the octupole deformation term. In Ref.\cite{Rad99} we have shown that in order to get the right deformation dependence of the single particle energies, $H_{mf}$ must be amended with a monopole-monopole interaction, $M\omega^2r^2\alpha_{00}Y_{00}$, where the monopole coordinate $\alpha_{00}$ is determined from the volume conservation restriction. This term has a constant contribution within a band. The constant value is, however, band dependent.

In order to find the eigenvalues of the model Hamiltonian we follow several steps:

1) In principle the single particle basis could be determined by diagonalizing $H_{mf}$ amended with the monopole interaction. The product basis for particle and core may be further used to find the eigenvalues of $H$. Due to some technical difficulties in restoring the
rotation and space reversal symmetries for the composite system wave function, this procedure is however tedious and therefore we prefer a simpler method.
Thus,  the single particle space consists of three spherical shell model states with angular momenta 
 $j_1,j_2, j_3$. We suppose that $j_1$ and $j_2$ have the parity $\pi=+$, while $j_3$ has a negative parity $\pi=-$. 
Due to the quadrupole-quadrupole interaction the odd particle from the state $j_1$ can be promoted to $j_2$ and vice-versa. The octupole-octupole interaction connects the states $j_1$ and $j_2$ with $j_3$. Due to the above mentioned effects, the spherical and space reversal symmetries of the single particle motion are broken. To be more specific, by diagonalizing $H$ (2.1) in a projected spherical particle-core basis with the spherical single particle state factors mentioned above, the eigenstates could be written as a projected spherical particle-core state having as  single particle state factor a function without good rotation and parity symmetries. Therefore, one could start with a coupled basis where the single particle state is a linear combination of the spherical states, where the mixing coefficients are to be determined by a least square fitting procedure as to obtain an optimal description of the experimental excitation energies.   Thus, instead of dealing with a spherical shell model state coupled to a deformed core without reflection symmetry, as the traditional particle-core approaches proceed, here the single particle orbits are lacking the spherical and space reversal symmetries and by this, their symmetry properties are consistent with those of the phenomenological core.        

2) We remark that  $\Psi_{g}$ is a sum of two states of different parities. This happens due to the specific structure of the octupole coherent state: 
\begin{equation}
\Psi_{o}=\Psi_{o}^{(+)}+\Psi_{o}^{(-)}.
\end{equation}
The states of a given angular momentum and positive parity  can be obtained through projection from the intrinsic states:
\begin{equation}
|n_1l_1j_1 K\rangle\Psi^{(+)}_{o}\Psi_{q},\;\;|n_2l_2j_2 K\rangle\Psi^{(+)}_{o}\Psi_{q},\;\;|n_3l_3j_3 K\rangle\Psi^{(-)}_{o}\Psi_{q}.
\end{equation}
The projected states of negative parity  are obtained from the states:
\begin{equation}
|n_1l_1j_1 K\rangle\Psi^{(-)}_{o}\Psi_{q},\;\;|n_2l_2j_2 K\rangle\Psi^{(-)}_{o}\Psi_{q},\;\;|n_3l_3j_3 K\rangle\Psi^{(+)}_{o}\Psi_{q}.
\end{equation}
The angular momentum and parity projected states are denoted by:
\begin{eqnarray}
\varphi^{(+)}_{IM}(j_iK;d,f)&=&N^{(+)}_{i;IK}P^{I}_{MK}|n_il_ij_iK\rangle \Psi^{(+)}_{o}\Psi_{q} \equiv N^{(+)}_{i;IK}\psi^{(+)}_{IM}(j_iK;d,f),i=1,2,\nonumber\\
\varphi^{(+)}_{IM}(j_3K;d,f)&=&N^{(+)}_{3;IK}P^{I}_{MK}|n_3l_3j_3K\rangle \Psi^{(-)}_{o}\Psi_{q}
\equiv N^{(+)}_{3;IK}\psi^{(+)}_{IM}(j_3K;d,f),\nonumber\\
\varphi^{(-)}_{IM}(j_iK;d,f)&=&N^{(-)}_{i;IK}P^{I}_{MK}|n_il_ij_iK\rangle \Psi^{(-)}_{o}\Psi_{q}\equiv N^{(-)}_{i;IK}\psi^{(-)}_{IM}(j_iK;d,f),i=1,2,\nonumber\\
\varphi^{(-)}_{IM}(j_3K;d,f)&=&N^{(-)}_{3;IK}P^{I}_{MK}|n_3l_3j_3K\rangle \Psi^{(+)}_{o}\Psi_{q}
\equiv N^{(-)}_{3;IK}\psi^{(-)}_{IM}(j_3K;d,f).
\end{eqnarray}
The factors $N^{(\pm)}_{i,IK}$ assure that the projected states $\varphi^{(\pm)}$ are normalized to unity. Obviously the unnormalized projected states are denoted by $\psi^{(\pm)}$.
For the quantum number $K$ we consider the lowest three values, i.e. $K=1/2,3/2,5/2$.
Note that the earlier particle-core approaches \cite{RaCea,Lea} restrict the single particle space to a single $j$, which results in
eliminating the contribution of the octupole-octupole interaction. 

3) Note that for a given  $j_i$, the projected states with different $K$ are not orthogonal.
Indeed, the overlap matrices :

\begin{eqnarray}
A^{(+)}_{K,K'}(Ij_l;d,f)&=&\langle \psi^{(+)}_{IM}(j_lK;d,f)|\psi^{(+)}_{IM}(j_lK';d,f)\rangle,\nonumber\\
                   l&=&1,2,3;\;K,K'=1/2,3/2,5/2,
\nonumber\\
A^{(-)}_{K,K'}(Ij_l;d,f)&=&\langle \psi^{(-)}_{IM}(j_lK;d,f)|\psi^{(-)}_{IM}(j_lK';d,f)\rangle ,
\nonumber\\
                  l&=&1,2,3;\;K,K'=1/2,3/2,5/2,
\end{eqnarray} 
are not diagonal.
By diagonalization, one obtains the eigenvalues $a^{(\pm)}_{Ip}(j_l)$ and the corresponding eigenvectors $V^{(\pm)}_{IK}(j_l,p)$, with $K=1/2,3/2,5/2$ 
and $p=1,2,3$.
Then, the functions:

\begin{eqnarray}
\Psi^{(+)}_{IM}(j_l,p;d,f)=N^{(+)}_{l;Ip}\sum_{K}V^{(+)}_{IK}(j_l,p)\psi^{(+)}_{IM}(j_lK;d,f),
\nonumber\\
\Psi^{(-)}_{IM}(j_l,p;d,f)=N^{(-)}_{l;Ip}\sum_{K}V^{(-)}_{IK}(j_l,p)\psi^{(-)}_{IM}(j_lK;d,f),
\label{psiplmi}
\end{eqnarray}
are mutually orthogonal. The norms are given by:
\begin{equation}
\left(N^{(\pm)}_{l;Ip}\right)^{-1}=\sqrt{a^{(\pm)}_{Ip}(j_l)}.
\end{equation}
For each of the new states, there is a term in the defining sum (\ref{psiplmi}), which has a maximal weight. The corresponding  quantum number $K$ is conventionally  assigned to the mixed state. 

4) In order to simulate the core deformation effect on the single particle motion, in some cases the projected states corresponding to different $j$ must be mixed up.

\begin{eqnarray}
\Phi^{(+)}_{IM}(p;d,f)=\sum_{l=1,2,3}{\cal A}^{(+)}_{pl}\Psi^{(+)}_{IM}(j_lp;d,f),\nonumber\\
\Phi^{(-)}_{IM}(p;d,f)=\sum_{l=1,2,3}{\cal A}^{(-)}_{pl}\Psi^{(-)}_{IM}(j_lp;d,f).
\label{Psiplmi}
\end{eqnarray}
The amplitudes ${\cal A}^{(\pm)}_{pl}$ can be obtained either by diagonalizing
 $H_{mf}$ or, as we mentioned before, by a least square fitting procedure applied to the excitation energies.
	
The energies of the odd system are approximated by the average values of the model Hamiltonian corresponding to the projected states:
\begin{eqnarray}
E^{(+)}_I(p;d,f)=\langle \Phi^{(+)}_{IM}(p;d,f)|H|\Phi^{(+)}_{IM}(p;d,f)\rangle,\nonumber\\
E^{(-)}_I(p;d,f)=\langle \Phi^{(-)}_{IM}(p;d,f)|H|\Phi^{(-)}_{IM}(p;d,f)\rangle.
\label{energ}
\end{eqnarray}

The matrix elements involved in the above equations can be analytically calculated.
Note that due to the structure of the particle-core projected states, the energies for the odd system are determined by the coupling of the odd particle to the excited states of the core ground band. 

The approach presented in this section was applied for the description of the $K^{\pi}=1/2^{\pm}$ bands.
However, this procedure can  be extended by including the $K\ne 0$ states in the space describing the deformed core.
\renewcommand{\theequation}{3.\arabic{equation}}
\setcounter{equation}{0}
\section{Description of the $K^{\pi}=\frac{3}{2}^{\pm}, \frac{5}{2}^{\pm}$ bands.}
In principle the method presented in the previous section may work for the description of bands with the quantum number $K$ larger than $1/2$. However the intrinsic reference frame for the odd system is determined by the deformed core and therefore one expects that this brings an important contribution to the quantum number $K$. To be more specific, we cannot expect that projecting out the good angular momentum from $|j,5/2\rangle \otimes \Psi_g$, a realistic description of the $K=5/2$ bands is obtained. Therefore we assume that the $K^{\pi}=\frac{3}{2}^{\pm}, \frac{5}{2}^{\pm}$ bands
are described by projecting out the angular momentum from a product state of a low $K$ single particle state and the intrinsic gamma band state. 

We recall that within CSM, the states of the gamma band are obtained by projection from the intrinsic state:
\begin{equation}
\Psi^{(\gamma;\pm)}_{2}=\Omega^{\dagger}_{\gamma,2}\Psi^{(\pm)}_{o}\Psi_{q},
\label{Psiga}
\end{equation}
where the excitation operator for the gamma intrinsic state is defined as:
\begin{equation}
\Omega^{\dagger}_{\gamma,2}=\left(b^{\dagger}_2b^{\dagger}_2\right)_{22}+d\sqrt{\frac{2}{7}}b^{\dagger}_{22}.
\end{equation}
The low index of $\Psi$ in Eq. (\ref{Psiga}) is the the $K$ quantum number for the $\gamma$ intrinsic state.
Thus, a simultaneous description of the bands with $K=1/2, 3/2, 5/2$ can be achieved with the projected states:
\begin{eqnarray}
\varphi^{(\pm)}_{IM;1/2}&=&N^{(\pm)}_{I,1/2}\sum_{J}\left(N^{(g,\pm)}_{J}\right)^{-1}C^{j_1~J~I}_{1/2\;0\;1/2}\left[|n_1l_1j_1\rangle\otimes\varphi^{(g;\pm)}_J\right]_{IM},\nonumber\\
\varphi^{(\pm)}_{IM;3/2}&=&N^{(\pm)}_{I,3/2}\sum_{J}\left(N^{(\gamma,\pm)}_{J}\right)^{-1}C^{\;\;j_2~\;\;J~\;I}_{-1/2\;\;2 \;3/2}\left[|n_2l_2j_2\rangle\otimes\varphi^{(\gamma;\pm)}_J\right]_{IM},\nonumber\\
\varphi^{(\pm)}_{IM;5/2}&=&N^{(\pm)}_{I,5/2}\sum_{J}\left(N^{(\gamma,\pm)}_{J}\right)^{-1}C^{j_3~J~I}_{1/2\;2\;5/2}\left[|n_3l_3j_3\rangle\otimes\varphi^{(\gamma;\mp)}_J\right]_{IM}.
\end{eqnarray}
In the above expressions the notation $N^{(i,\pm)}_J$ with $i=g,\gamma$ is used for the normalization factors of the projected states describing the ground and the gamma bands, respectively, of the even-even core.
Note that for each angular momentum $I$ the above set of three projected states is orthogonal.

The energies for the six  bands with $K^{\pi}=1/2^{\pm},3/2^{\pm},5/2^{\pm}$ are obtained by
  averaging  the model Hamiltonian (\ref{Hamodd}) with the projected states defined above. 
\begin{equation}
E_{I,K}=\langle \varphi^{(\pm)}_{IM;K}|H|\varphi^{(\pm)}_{IM;K}\rangle, K=1/2, 3/2, 5/2.
\end{equation}

The matrix elements of the particle-core interaction are given in Appendix A
\renewcommand{\theequation}{4.\arabic{equation}}
\setcounter{equation}{0}

\section{Transition probabilities}
For some $K=1/2$ bands, results for the reduced $E1$ and $E2$ transition probabilities are available. They are given in terms of the branching ratios:
\begin{equation}
R_{I^{\pi}}= \frac{B(E1;I^{\pi}\to (I-1)^{\pi^{'}})}{B(E2;I^{\pi}\to (I-2)^{\pi})},\pi^{'}\ne \pi .
\label{branch}
\end{equation}
To describe these data we use the wave functions defined in Section II. We recall that the positive parity states are obtained by coupling the spherical shell model state $j_1$ or $j_2$ to a positive
parity core with a small admixture of a state coupling $j_3$ and a negative parity core. On the other hand the negative parity states are given by coupling one of the states $j_1$ or $j_2$ to a negative parity core and a small component consisting in a product state of $j_3$ and a positive parity core-state. Thus, the single particle E1 transition operator may connect the leading term of the initial state with the small component of the final state. One expects that the contribution of this term to the E1 transition is negligible comparing it with the contribution of collective dipole operator. Therefore the dipole transition operator considered in the present paper  is the boson operator:
\begin{equation}
Q_{1\mu}=eq_{1}\left((b^{\dagger}_2b^{\dagger}_3)_{1\mu}+(b_3b_2)_{\widetilde{1\mu}}\right).
\end{equation}
Concerning the quadrupole transition operator, this has the structure:
\begin{equation}
Q_{2\mu}=eq_2\left(b^{\dagger}_{2\mu}+(-)^{\mu}b_{2,-\mu}+ar^2Y_{2\mu}\right).
\end{equation}
The branching ratio (\ref{branch}) for the initial state $I^{\pi}$ is:
\begin{equation}
R_{I^{\pi}}=\left[\frac{\langle I^{\pi}||Q_{1}||(I-1)^{\pi^{'}}\rangle}
{\langle I^{\pi}||Q_{2}||(I-2)^{\pi}\rangle}\right]^2 .
\end{equation}
Here the initial and final states are mixture of different $K$ states as well as mixture of the $j$ states defined by 
Eq.(\ref{Psiplmi}). The matrix elements of the transition operators between the basis states are given in Appendix B$^1$
\setcounter{footnote}{1}\footnotetext{Throughout this paper the reduced matrix elements are defined according to Rose's convention \cite{Rose}.}.

\renewcommand{\theequation}{5.\arabic{equation}}
\setcounter{equation}{0}

\section{Numerical results}
The results obtained in Section II have been used to calculate the excitation energies for one positive and one negative parity bands in three even-odd isotopes: $^{219}$Ra, $^{237}$U and $^{239}$Pu. The parameters defining $H_{core}$, as well as the deformation parameters $d$ and $f$ are those used to describe the properties of  eight rotational bands in the even-even neighboring isotopes. The single particle states are spherical shell model states with the appropriate parameters for the $(N,Z)$ region of the considered isotopes \cite{Ring}. Our calculations for the mentioned odd isotopes correspond to the single particle states: $(j_1,j_2,j_3)=(2g_{7/2},2g_{9/2},1h_{9/2})$. 
In order to obtain the best agreement between the calculated excitation energies and the corresponding experimental data, in the expansion (\ref{Psiplmi}) a small admixture of the states $(j_1;j_3)$ and $(j_2;j_3)$ was considered: $|{\cal A}^{(+)}_{i,3}|^2$ and $|{\cal A}^{(-)}_{i,3}|^2$,  are both equal to 0.001 for for $^{219}$Ra, while for $^{237}$U and $^{239}$Pu the amplitudes take the common value: $|A^{(+)}_{i,3}|^2= |A^{(-)}_{1,3}|^2=0.04$.   The mixing amplitude of the states $(j_1,j_2)$ is negligible small. Energies
(\ref{energ}) depend on the interaction strengths $X_2,X_3, X_{jJ}$ and $X_{I^2}$.
These were determined by fitting four particular energies in the two bands of different parities, i.e. $K^{\pi}=\frac{1}{2}^{\pm}$. The results of the fitting procedure are given in Table I. Inserting these in Eqs. (\ref{energ}), the energies in the two bands with $K=1/2$ are readily obtained.
\begin{equation}
E(I^{\pm})=E^{(\pm)}_{I}(1;d,f)-E^{(+)}_{\frac{1}{2}}(1;d,f).
\end{equation}

 The theoretical results for excitation energies, listed in Tables II and III, agree quite well with the corresponding experimental data. The levels for the three isotopes have been populated by different experiments. Indeed, the $K^{\pi}=1/2^{\pm}$ bands have been identified in $^{219}$Ra with  the reaction   $^{208}$Pb($^{14}$C,3n)$^{219}$Ra \cite{Cottle}, in $^{237}$U via a pickup reaction on a $^{238}$U target, while in $^{239}$Pu with the so-called "unsafe" Coulomb excitation technique \cite{Zhu}.
 Our results suggest that the dominant $K$ component is $K=1/2$ while the dominant $j$ component is $g_{9/2}$.
To have a measure for the agreement quality, we calculated the r.m.s. values for the deviations of the calculated values from the experimental ones. The results for $^{219}$Ra, $^{237}$U and $^{239}$Pu are 66.24 keV, 48.97 keV and 31.8 keV, respectively. In calculating the $r.m.s.$ value for $^{219}$Ra we ignored the data for the states $53/2^{\pm}$ since the spin assignment is unsure. It is interesting to mention that the spectrum of $^{219}$Ra has been measured by two groups \cite{Cottle,Wiel}
by the same reaction, $^{208}$Pb($^{14}$C,3n)$^{219}$Ra. However they assign for the ground state different angular momenta,
$9/2^+$ \cite{Cottle} and $7/2^+$\cite{Wiel}. In our approach both assignments yield good description of the data. However we made the option for $9/2^+$ since the corresponding results agree better with the experimental data than those obtained with the other option. The results and experimental data for $^{219}$Ra are plotted in Fig.1.

The case of $^{227}$Ra was treated with the formalism presented in Section III. The single particle basis is:
$2g_{7/2}, 2g_{9/2}, 2f_{5/2}$. The first state coupled to the coherent state describing the unprojected ground state, i.e.
$2g_{7/1,1/2}\Psi_g$, 
generates the parity partner bands $K^{\pi}=1/2^{\pm}$. The bands $K^{\pi}=3/2^{\pm}$ are obtained through projection from
the product state $2g_{9/2, -1/2}\Psi^{(\gamma;\pm)}_2$ while the bands  $K^{\pi}=5/2^{\pm}$ originate from the intrinsic state 
$2f_{5/2,1/2}\Psi^{(\gamma;\mp)}_2$. Concerning the bands characterized by $K^{\pi}=1/2^{\pm}$ one could consider also the mixing of components with different K in the manner discussed in Section II. However, our numerical application suggests that such a mixing is not really necessary in order to obtain a realistic description of the available data. The calculated energies in the three bands are compared with the corresponding experimental data in Fig.2. The plotted values are collected in Table IV. The states for $^{227}$Ra have been obtained  in Ref.\cite{Egidy} by using the $(n,\gamma), (d,p)$ and $(\vec{t},d)$ reactions and the $\beta^-$ decay of $^{227}$Fr. The spectrum yielded by the mentioned experiments was interpreted in Ref.
\cite{LeaChen} in terms of a particle-core interaction.

\begin{table}
\begin{tabular}{|c|cccc|}
\hline
Parameters     &$^{219}$Ra &$^{227}$Ra        &   $^{237}$U      &   $^{239}$Pu\\
\hline                                   
$X_2b^2$[keV] &22.714&-1.992        &1.080            &-2.515 \\
$X_3b^3$[keV] &-8.823&169.511        &2.227           &4.937\\
$X_{jJ}$[keV] &-0.230&8.553     &-5.817           &-3.985 \\
$X_{I^2}$[keV]&3.778 &4.390    &4.634            &5.050  \\
\hline
\end{tabular}
\caption{Parameters involved in the particle-core Hamiltonian obtained by fitting four excitation energies. Here $b$ denotes
the oscillator length: $b=(\frac{\hbar}{M\omega})^{1/2};\,\hbar\omega =41A^{-1/3}$. The usual notations for nucleon mass (M) and atomic number (A) were used.}
\end{table}

\begin{table}
\begin{tabular}{|c|cc|cc|}
\hline
    &\multicolumn{4}{c|}{$^{219}$Ra}\\
\hline
    & \multicolumn{2}{c|}{$\pi =+$}&\multicolumn{2}{c|}{$\pi= -$}\\
    \hline
J     &    Exp.    &   Th.          &   Exp.     &      Th. \\
9/2   &   0.0      & 0.0            &            &           \\ 
13/2  &  234.3     &  235.4         &            &           \\
15/2  &            &                &   495.4    & 496.0           \\
17/2  &  529.1     &  526.7         &            &            \\
19/2  &            &                &   733.7    &  729.1          \\
21/2  &  876.6     & 863.4          &            &            \\
23/2  &            &                &  1035.6    &  1029.0           \\
25/2  &  1271.6    & 1238.1         &            &             \\
27/2  &            &                &  1393.6    &  1388.4           \\ 
29/2  &  1684.7    & 1646.8         &            &             \\
31/2  &            &                &  1815.6    &  1800.2           \\
33/2  &  2113.4    & 2088.4         &            &             \\
35/2  &            &                &  2272.1    &  2258.2           \\
37/2  &  2563.6    & 2564.2         &            &             \\
39/2  &            &                &  2750.8    &  2756.7            \\
41/2  &  3029.0    & 3076.5         &            &              \\
43/2  &            &                &  3255.8    &  3291.6            \\
45/2  &  3505.0    & 3627.8         &            &              \\
47/2  &            &                &  3776.5    &  3859.8            \\
49/2  &  4009.6    & 4219.9         &            &              \\
51/2  &            &                &  4328.9    &  4459.6            \\
53/2  &  4540.4    & 4784.7         &            &              \\
55/2  &            &                &  4913.6    &  5078.5            \\
\hline
\end{tabular}
\caption{Excitation energies in
$^{219}$Ra   for the bands characterized by
 $K^{\pi}=\frac{1}{2}^{+}$ and  $K^{\pi}=\frac{1}{2}^{-}$ respectively,   
 are given in keV. The results of our calculations (Th.) are compared with the corresponding experimental data (Exp.) taken from Ref.\cite{Cottle}.}
\end{table}

\begin{table}
\begin{tabular}{|c|cc|cc|cc|cc|}
\hline
    &\multicolumn{4}{c|}{$^{237}$U}&\multicolumn{4}{c|}{$^{239}$Pu}\\
\hline
    & \multicolumn{2}{c|}{$\pi =+$}&\multicolumn{2}{c|}{$\pi= -$}
& \multicolumn{2}{c|}{$\pi =+$}&\multicolumn{2}{c|}{$\pi= -$}\\
    \hline
J     &    Exp.    &   Th          &   Exp.     &      Th. &
Exp. & Th. & Exp. & Th. \\ 
\hline
1/2  &0.0         &  0.0          &            &398.5
&0.0            &0.0          &469.8         &469.8\\
3/2  &11.4        &11.4           &            &454.4
&7.9            &7.9          &492.1         &477.7\\
5/2  &56.3        &74.6           &            &475.5
&57.3           &62.8         &505.6         &498.3\\
7/2  &82.9        &106.9          &            &550.3
&75.7           &108.4        &556.0         &549.8\\
9/2  &162.3       &191.2          &            &581.3
&163.8          &183.5        &583.0         &572.0\\
11/2 &204.1       &231.8          &            &680.9
&193.5          &222.0        &661.2         &655.2\\
13/2 &317.3       &347.7          &            &721.9
&318.5          &338.1        &698.7         &685.7\\
15/2 &375.1       &393.1          &846.4       &846.4
&359.2          &386.5        &806.4         &799.9\\
17/2 &518.2       &544.2          &930.0       &899.1
&519.5          &534.9        &857.5         &839.5\\
19/2 &592.0       &592.0          &1027.5      &1046.6
&570.9          &592.2        &992.5         &984.2\\
21/2 &762.8       &780.3          &1131.0      &1113.3
&764.7          &773.7        &1058.1        &1033.3\\
23/2 &853.0       &829.0          &1250.7      &1281.3
&828.0          &839.2        &1219.4        &1208.3\\
25/2 &1048.7      &1065.8         &1376.1      &1364.8
&1053.1         &1054.4       &1300.9        &1267.2\\
27/2 &1155.1      &1108.8         &1515.7      &1550.2
&1127.8         &1127.8       &1487.4        &1472.2\\
29/2 &1372.2      &1378.3         &1662.3      & 1654.0
&1381.5         &1377.0       &1584.9        &1541.2\\
31/2 &1494.1      &1421.6         &1821.8      &1852.8
 &1467.8         &1458.0       &1795.4        &1776.0\\
33/2 &1729.2      &1728.7         &1987.7      &1981.0
&1748.5         &1744.2       &1908.9        &1855.4\\
35/2 &1868.2      &1772.5         &2166.5      &2188.9
&1847.0         &1831.3       &2143.4        &2119.8\\
37/2 &2117.2      &2117.2         &2349.7      &2346.1
 &2152.2         &2150.2       &2272.0        &2209.8\\
39/2 &2272.2      &2161.7         &2547.5      &2558.3
&2263.0         &2245.0       &2529.4        &2503.6\\
41/2 &2530.1      &2544.1         &2746.7      &2749.4
 &2590.1         &2597.9       &2672.0        &2604.4\\
43/2 &2702.5      &2589.4         &2960.5      &2960.5
&2714.0         &2700.5       &2951.4        &2927.5\\
45/2 &2963.8      &3009.5         &3174.7      &3191.3
&3060.1         &3087.5       &3108.0        &3039.3\\
47/2 &3154.5      &3055.6         &3401.5      &3395.3
 &3198.0         &3198.0       &3407.0        &3395.3\\
49/2 &3415.8      &3513.7         &3630.0      &3671.7
 &3559.1         &3619.1       &3578.0        &3514.4\\
51/2 &3625.5      &3560.5         &3865.0      &3862.4
 &3713.0         &3737.0       &3895.0        &3895.8\\
53/2 &3886.8      &4057.8         &4105.0      &4190.9
&4087.1         &4194.0       &4080.0        &4029.9\\
55/2 &4115.0      &4104.8         &4344.0      &4350.0
&4256.0         &4319.8       &4413.0        &4436.7\\
\hline
\end{tabular}
\caption{Excitation energies in
$^{237}$U  and $^{239}$Pu, for the bands characterized by $K^{\pi}=\frac{1}{2}^{+}$ and  $K^{\pi}=\frac{1}{2}^{-}$ respectively,   
 are given in keV. The results of our calculations (Th.) are compared with the corresponding experimental data (Exp.) taken from Ref.\cite{Zhu}.}
\end{table}

\begin{figure}[ht!]
\begin{center}
\includegraphics[width=0.9\textwidth]{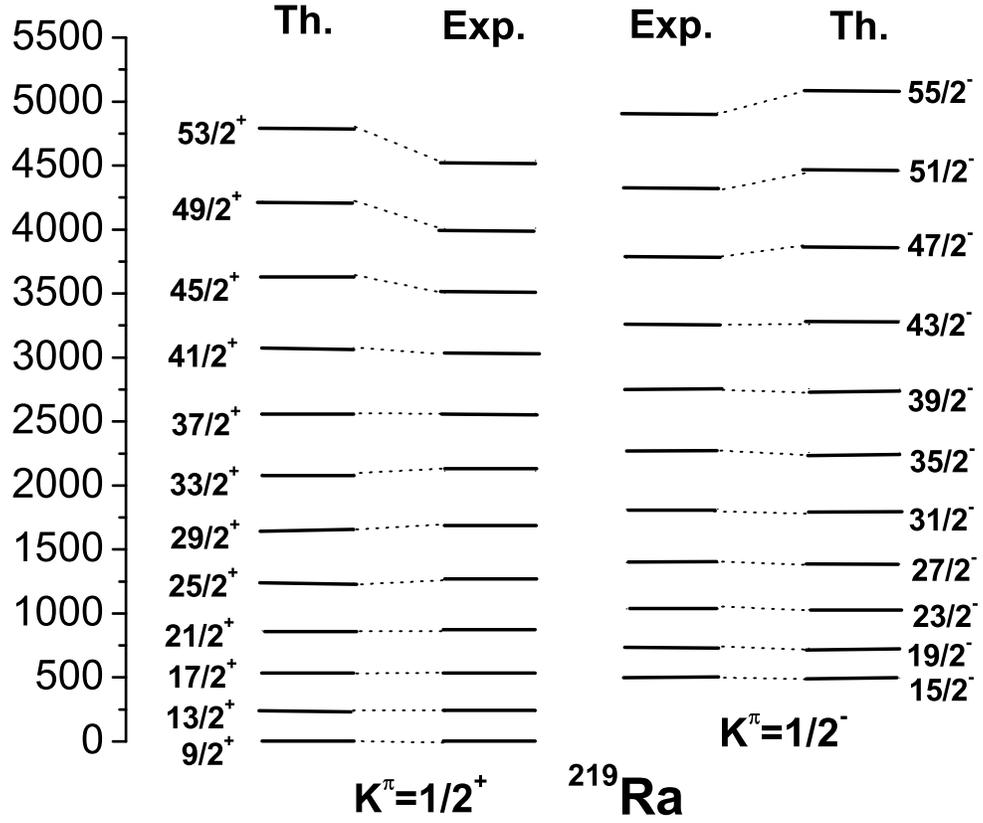}
\end{center}
\caption{Calculated (Th.) and experimental (Exp.) excitation energies 
for the $K^{\pi}=\frac{1}{2}^{\pm}$ bands in $^{219}$Ra. The data were taken from Ref.\cite{Cottle}.}
\label{Fig. 1}
\end{figure}

\begin{figure}[ht!]
\begin{center}
\includegraphics[width=0.9\textwidth]{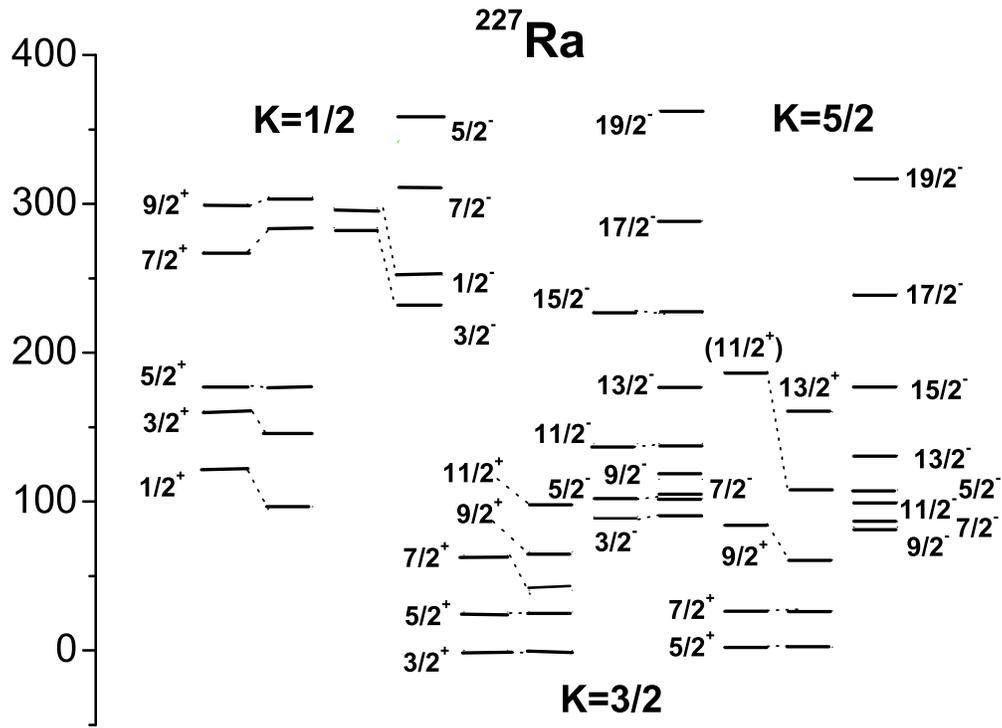}
\end{center}
\caption{ Calculated and experimental excitation energies for the bands with
$K^{\pi}=\frac{1}{2}^{\pm},\frac{3}{2}^{\pm},\frac{5}{2}^{\pm} $ in $^{227}$Ra.
The experimental data were taken from Ref.\cite{Egidy}.}
\label{Fig. 2}
\end{figure}

\begin{table}
\begin{tabular}{|c|cc|cc|cc|cc|cc|cc|}
\hline
    &\multicolumn{12}{c|}{$^{227}$Ra}\\
\hline
    &\multicolumn{4}{c|}{K=1/2}&\multicolumn{4}{c|}{K=3/2}&\multicolumn{4}{c|}{K=5/2}\\
\hline
 J   & \multicolumn{2}{c|}{$\pi =+$}&\multicolumn{2}{c|}{$\pi= -$}
    & \multicolumn{2}{c|}{$\pi =+$}&\multicolumn{2}{c|}{$\pi= -$}
    & \multicolumn{2}{c|}{$\pi =+$}&\multicolumn{2}{c|}{$\pi= -$}\\
\hline
      &Exp.&Th.     & Exp.& Th.    & Exp.&Th.&  Exp.&Th.&Exp.&Th. &Exp.&Th.\\
1/2  &121& 96.6     &297&251.8    &     &     &    &    &   &   & $\hskip0.6cm$   &  \\
3/2  &161& 145.5    &284& 232.4   & 0.0 &0.0  &90  & 90 &   &       &     &   \\
5/2  &177& 177.0    &    &359.1   & 26  &26.0 &102 & 102& 2 & 2.    &    & 107.6 \\ 
7/2  &268& 283.6    &    &310.6   & 64  &40.33 &   &104.6&26.&26.5  &    & 86.6 \\
9/2  &300& 304.6    &    &        &     &66.2 &    &115.1&84 &61.0  &    & 82.8  \\
11/2 &    &574.5    &    &        &     &97.9 &139 &139.1&187& 107.5&    & 99.9  \\
13/2 &    &         &    &        &     &140.5&    &176.9&   & 160.1&    & 131.1  \\
15/2 &    &         &    &        &     &     &228 &226.6&   & 221.0&    & 177.5\\
17/2 &    &         &    &        &     &     &    &288.4&   & 291.4&    & 239.6  \\
19/2 &    &         &    &        &     &     &    &     &   & 372.3&    & 317.6 \\
\hline
\end{tabular}
\caption{Excitation energies in
$^{227}$Ra   for the bands characterized by
 $K^{\pi}=\frac{1}{2}^{\pm}, \frac{3}{2}^{\pm}, \frac{5}{2}^{\pm}$ respectively,   
 are given in keV. The results of our calculations (Th.) are compared with the corresponding experimental data (Exp.) taken from Ref.\cite{Egidy}.}
\end{table}

\begin{figure}[ht!]
\begin{center}
\includegraphics[width=0.8\textwidth]{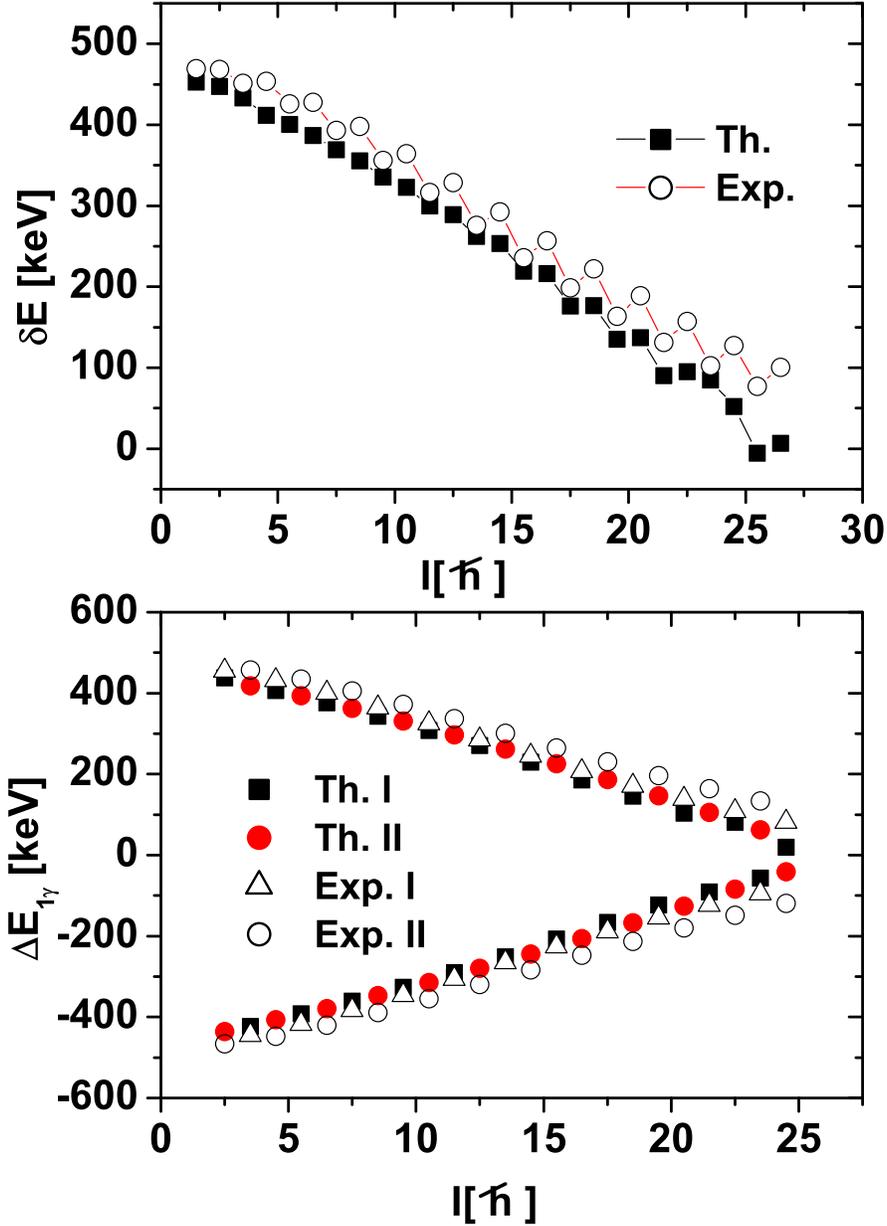}
\end{center}
\caption{The theoretical and experimental energy displacement functions $\delta E(I)$ and $\Delta E_{1,\gamma}(I)$ given by Eqs.(\ref{delta}) and (\ref{Delta}) respectively, characterizing  the isotope $^{239}$Pu, are plotted as a function of the angular momentum $I$.
Experimental data are taken from Ref.\cite{Zhu}. In the lower panel, the theoretical and experimental  
$\Delta E_{1,\gamma}(I)$ corresponding to the states $I^{\pi}=\left(\frac{1}{2}+2k\right)^+$ with k=1,2,3,..., are represented by the symbols labeled by
$Th.I$ and $Exp.I$  respectively, while those associated with the negative parity states $I^{\pi}=\left(\frac{1}{2}+2k\right)^-$ with k=1,2,3,... bear the labels  $Th.II$ and $Exp. II$
, respectively.}
\label{Fig. 3}
\end{figure}
\clearpage

\begin{table}
\begin{tabular}{|c|ccc|}
\hline
    &\multicolumn{3}{c|}{$\frac{B(E1;J\to (J-1))}{B(E2;J\to (J-2))}[10^{-6}fm^{-2}]$}\\
\hline
 $J^{\pi}-J_{g.s.}$       &   Exp.  &  present& Ref.\cite{Zub1}   \\
\hline
5$^-$  &2.52(18) & 2.52& 1.195    \\      
6$^+$  &1.12(08) & 1.09& 0.314    \\
7$^-$  &1.49(10) & 3.97& 1.318    \\
8$^+$  &1.23(16) & 1.23& 0.313    \\
9$^-$  &1.16(08) & 4.56& 1.442    \\
10$^+$ &2.77(64) & 1.44& 0.312    \\
11$^-$ &1.41(9)  & 4.59& 1.567    \\
12$^+$ &3.68(26) & 1.69& 0.313    \\
13$^-$ &2.14(30) & 4.39& 1.691    \\
14$^+$ &1.96(14) & 1.96& 0.314    \\
15$^-$ &1.76(18) & 4.11& 1.814    \\
16$^+$ &1.06(17) & 2.22& 0.315    \\
17$^-$ &2.08(28) & 3.84& 1.936    \\
18$^+$ &3.34(48) & 2.45& 0.317    \\
19$^-$ &1.34(42) & 3.62& 2.057    \\
20$^+$ &2.38(44) & 2.63& 0.318    \\
21$^-$ &4.01(94) & 3.44& 2.177    \\
Average&2.09(9)  & 2.97& 1.072    \\
\hline
\end{tabular}
\caption{The experimental (Exp.) and calculated (present)  ratios $B(E1)/B(E2)$ for initial state
$J^{\pi}$ running from $19/2^-$ up to $51/2^-$. As mentioned in the text, $J_{g.s.}=9/2$. Experimental data are from Ref.\cite{Cottle}. The results are given in units of $10^{-6}fm^{-2}$. For comparison the results of Ref.\cite{Zub1} are also given in the column 3.}
\end{table}

From Fig. 2 we note that our approach reproduces the experimental energies ordering in the band $K^{\pi}=1/2^-$.
The energy split of the states $3/2^-, 1/2^-$ is nicely described although the doublet is shifted down by an amount of about
50 keV. In the band $5/2^+$ there exists an energy level which is tentatively assigned with $11/2^+$.
Our calculations suggests that this level could be assigned as $13/2^+$. No experimental data are available for the band $5/2^-$. In Fig. 2 we gave however the results of our calculations for this band. Note that the ordering for the lowest levels is not the natural one. However starting with $13/2^-$ the normal ordering is restored. It is interesting to note that the heading states for the bands $1/2^+$ and $5/2^+$ are almost degenerate. The same it is true for the lowest angular momenta states in their negative parity partner bands. The deviations r.m.s. for this nucleus is 23 keV. 

Now we would like to comment on the parameters yielded by the fitting procedure, for the considered isotopes.
Except for $^{237}$U, where both quadrupole-quadrupole and octupole-octupole interactions are attractive, 
the two interactions have different characters for the rest of nuclei. In the first situation the $\lambda$ (=2,3)-pole moments of  the odd nucleon and that of the collective core have different signs. In the remaining cases the two moments are of similar sign. We also remark the large strength for the $q_3Q_3$ interaction in $^{227}$Ra which is consistent with the fact that the neighboring even-even isotope exhibits a relatively large octupole deformation. Indeed, according to Ref.\cite{Rad006} for this nucleus we have $f=0.8$. The large value of the strength $X_3$ determines  large mixing amplitudes of the states
$[g_{9/2}\Psi^{(+)}_g; f_{5/2}\Psi^{(-)}_g]$ as well as of the states $[g_{9/2}\Psi^{(-)}_g; f_{5/2}\Psi^{(+)}_g]$.
Indeed, the value obtained for this amplitude is:  $|A^{(+)}_{i,3}|^2= |A^{(-)}_{1,3}|^2=0.07425$.
Another distinctive feature for $^{227}$Ra consists in the fact that the $jJ$ interaction strength has a  sign which is different from that associated to other nuclei. In fact the repulsive character of this interaction in $^{227}$Ra is necessary in order to compensate the large attractive contribution of the $q_3Q_3$ interaction.    
  
Further, we addressed the question whether one could identify signatures for static octupole deformation in the two bands.
To this goal, in Fig. 3, we plotted the energy displacement functions \cite{Rad02,Rad03,Bona0}:
\begin{eqnarray}
\delta E(I)&=&E(I^-)-\frac{(I+1)E((I-1)^+)+IE((I+1)^+)}{2I+1},\nonumber\\
\label{delta}\\
\Delta E_{1,\gamma}(I)&=&\frac{1}{16}[6E_{1,\gamma}(I)-4E_{1,\gamma}(I-1)-4E_{1,\gamma}(I+1)\nonumber\\
&&+E_{1,\gamma}(I-2)+E_{1,\gamma}(I+2)],\label{Delta}\\
E_{1,\gamma}(I)&=&E(I+1)-E(I).\nonumber
\end{eqnarray}
The first function, $\delta E$, vanishes when the   excitation energies of the parity partner bands depend linearly on
$I(I+1)$ and, moreover, the moments of inertia of the two bands are equal. Thus, the vanishing value of $\delta E$ is considered to be a signature for octupole deformation. If the excitation energies depend quadratically on $I(I+1)$ and the coefficients of the $[I(I+1)]^2$ terms for the positive and negative parity bands are equal, the second energy displacement function $\Delta E_{1,\gamma}$ vanishes, which again suggests that a static octupole deformation shows up. The parities associated to the angular momenta, involved in $\Delta E_{1,\gamma}$ 
are as follows: the levels $I$ and $I\pm 2$ have the same parity, while levels $I$ and $I\pm 1$ are of opposite parities.
The results plotted in Fig. 3 correspond to $^{239}$Pu. We choose this nucleus, since more data are available. The plot suggests that a static octupole deformation is possible for the states with angular momenta $I\ge \frac{51}{2}$ belonging to the two parity partner bands.

Finally we calculated the branching ratio $R_J$ defined by Eq.(\ref{branch}), for $^{219}$Ra. There are two parameters involved which were fixed so that two particular experimental data are reproduced. The values obtained for these parameters are:
\begin{equation}
\frac{q_1}{q_2}=18.377\times 10^{-3}fm^{-1},\;\;ab^2=-0.63616 fm^2
\end{equation}
where $b$ denotes the oscillator length characterizing the spherical shell model states for the odd nucleon. 
As shown in Table V, the theoretical results agree reasonably well to the corresponding experimental data. Our results show an oscillating behavior with maxima for the negative parity states. Note that some off the data are well described while others deviate from the data by a factor ranging from 2 to 3. In the third column of Table V we listed the results obtained in
Ref.\cite{Zub1} by a different model. In the quoted reference the ratios corresponding to positive parity states are almost constant and small.

The spectra of the odd isotopes, considered here, have been previously studied in Refs.\cite{Zub,Zub1,Deni,Bona} using a quadrupole-octupole Hamiltonian in the intrinsic  deformation variables $\beta_2$ and $\beta_3$ separated in a  kinetic energy, a potential energy term and a Coriolis interaction. Due to the specific structure of the model Hamiltonian, an analytical solution for the excitation energies in the two bands of opposite parities was possible. It was shown that the split of the parity partner bands is determined by a combined effect coming from the Coriolis interaction, which affects the $K=\frac{1}{2}$ bands, and a quantum number $k$ associated to the motion of a phase angle $\phi$,
characterizing both the quadrupole and the octupole deformation variables. Based on analytical calculations,  some conclusions concerning the $B(E2)$ values associated to the intraband transitions between states of similar parities, have been drawn. Thus, if the odd particle state is of positive parity, the transitions between positive parity states are enhanced with respect to those connecting negative parity states. If the parity of the odd particle state is negative the ordering of the mentioned transitions is reversed.

Comparison between the present formalism and that of Ref. \cite{Bona} reveals the following features:

a) Having in mind the asymptotic behavior of the coherent states written in the intrinsic frame of reference \cite{Rad81}, one may anticipate that the wave function describing the odd system from Ref. \cite{Bona}, might be recovered in  the asymptotic limit of the present approach. Due to the fact that our formalism is associated to the laboratory reference frame, the Coriolis interaction does not show up explicitly. The split of the states of different parities is determined by the matrix elements of $H_{pc}$. Indeed, the quadrupole-quadrupole interaction has different matrix elements in the space of positive parity states $\Phi^{(+)}_{IM}(p;d,f)$ than in the space spanned by
$\Phi^{(-)}_{IM}(p;d,f)$ (see (2.12)). Note that the octupole-octupole interaction does not connect the states $\Phi^{(+)}_{IM}(p;d,f)$
and $\Phi^{(-)}_{IM}(p;d,f)$ but the diagonal elements corresponding to the mentioned states are different since so are the mixing amplitudes ${\cal A}^{(+)}_{pl}$ and ${\cal A}^{(-)}_{pl}$. The {\it spin-orbit} like interaction is also very important in determining the band parity split. The set of  angular momenta $J$ characterizing the core system, which participate in building up the angular momentum $I^+$ is very different from that involved in the structure of the state $I^-$. Therefore, summing the quantity $\frac{1}{2}[I(I+1)-j(j+1)-J(J+1)]$ with different
weights for the parity partner states $I^+$ and $I^-$ and then comparing the results, one certainly obtains the energy split caused by the term $\vec{j}\cdot\vec{J}$. Concluding, one may say that we identified three distinct sources generating a split for the parity partner states in the even-odd nuclei. While in Ref.\cite{Bona} $K$ is a good quantum number here $K$ labels the leading component in an expansion characterizing a  wave function with a definite angular momentum and a definite angular momentum z-projection, in the laboratory frame. Thus, although one says that $K=\frac{1}{2}$ since the corresponding component in the above mentioned expansion prevails, the mixing of different $K$
 components due to the single particle mixed states as well as due to the core projected states is considered in a natural manner. Therefore, one expects that the complex structure of the model states might be suitable for the description of the  transition probabilities between states from the two bands.

b) Since the coherent states are axially symmetric functions (only the boson components $b^{\dagger}_{\lambda 0}$, $\lambda =2, 3,$ are used in Eq. (\ref{cohq2q3}) defining the coherent states) we don't account for the motion of the $\gamma$-like deformation. Again the two formalisms are on a par with each other.

c) In general, the quadrupole-octupole boson descriptions overestimate the contribution to the system energy coming from the rotational degrees of freedom, since the Euler angles defining the intrinsic reference frame are involved in both the quadrupole and octupole  bosons. This ambiguity is however removed in our approach due to the angular momentum projection operation. The description used in Ref.\cite{Bona} is also not confronted with such a difficulty. 

d) The approach of Ref. \cite{Bona} is of a strong coupling type and therefore $K$ is a good quantum number, which is not the case
in the present paper. Indeed, we use the laboratory frame and the meaning of the quantum number $K$ is given by the fact that the $K$-component of the spherical function prevails over the components with $K^{\prime}\ne K$.  

e) The Hamiltonian describing the odd system (\ref{Hamodd}) involves a term $H_{core}$ which describes in a realistic fashion the neighboring even-even system. Indeed, this has been used in Ref.\cite{Rad06} to describe simultaneously eight rotational bands, four of positive and four of negative parity. By contrast, in Ref.\cite{Bona} the terms associated to the core are not appropriate for describing the complex structure of the even-even sub-system.

f) The agreement obtained in our approach for $^{239}$Pu is better than that shown in Ref.\cite{Bona}. However, the results from Ref.\cite{Bona} for $^{237}$U  agree better, with the corresponding data, than ours. Indeed, the $r.m.s$. values for the deviations of theoretical results from experimental data, reported in Ref.\cite{Bona}, are 30 keV and 60 keV for $^{237}$U and $^{239}$Pu,  which are to be compared with 48.9 keV and 31.8 keV respectively, obtained with our approach.

g) For some isotopes, in Ref.\cite{Bona}, the bands with $K=5/2$ are  solely considered. By contradistinction we treated simultaneously the bands with $K=1/2, 3/2, 5/2$, respectively. Moreover, a distinctive feature is the fact that here the bands with $K=3/2$ and $K=5/2$ are generated by coupling a single particle state to the states belonging to the $\gamma$ band of the core system.

\renewcommand{\theequation}{4.\arabic{equation}}
\setcounter{equation}{0}

\section{Conclusions}
In the previous sections we proposed a new formalism for the description of parity partner bands in even-odd nuclei.
Our approach uses a particle-core Hamiltonian, with a phenomenological core described in terms of quadrupole and octupole bosons.
The single particle space consists of three spherical shell model states, two of them having positive parity while the third one  a negative parity. The particle-core coupling terms cause the excitation of the odd particle from one state to any of the remaining two. Thus, the particle-core interaction might break two symmetries for the single particle motion, the rotation and space reflection, which, as a matter of fact, is consistent with the structure of the mean field obtained by averaging the model Hamiltonian with a quadrupole and octupole boson coherent state. For $K=1/2$ bands the single particle states are coupled to the ground state of a deformed core while for $K=3/2,5/2$ the single particle states are coupled to the gamma intrinsic state. The bands are generated through angular momentum projection from the particle-core intrinsic states mentioned above.
In this way the influence of the excited states from the ground band on the structure of the $K^{\pi}=1/2^{\pm}$ and that of the excited states from the $\gamma$ band on the $K^{\pi}=3,2^{\pm},5/2^{\pm}$ bands are taken into account. The contribution of various terms of the model Hamiltonian are analyzed in terms of the magnitude and the sign of the interaction strengths
yielded by the fitting procedure. Approaches which treat the particle-core interaction in the intrinsic frame of reference 
stress on the role played by the  Coriolis interaction, through the decoupling parameter, in determining the energy splitting of the parity partner states with $K=1/2$. For example, in $^{227}$Ra the decoupling factor is quite high (0.7)
\cite{Egidy}. In the laboratory frame we identified the interaction which determine the energy parity split.   

Application to $^{219}$Ra, $^{237}$U and $^{239}$Pu shows a good agreement between the calculated excitation energies in the bands with $K^{\pi}=\frac{1}{2}^{\pm}$ and the corresponding experimental data. The branching ratios of $^{219}$Ra have been also calculated. The agreement with the available data is quite good. Finally the results for a simultaneous treatment of 
six bands, $K^{\pi}=1/2^{\pm},3/2^{\pm},5/2^{\pm}$, were presented for $^{227}$Ra.  The plot for the energy displacement functions, or energy staggering factors, made for $^{239}$Pu, indicates that a static octupole deformation might be set on for states with angular momentum larger than $\frac{51}{2}\hbar$.

Before closing, we would like to add few remarks about the possible development of the present formalism. Choosing for the core unprojected states, the generating states for the parity partner bands with $K^{\pi}=0_{\beta}^{\pm}, 1^{\pm}$ states, otherwise keeping the same single particle basis for the odd nucleon, the present formalism can be extended to another four bands, two of positive and two of negative parity. Another noteworthy remark refers to the chiral symmetry \cite{Frau} for the composite particle and core system. Indeed, in Ref.\cite{Rad006} we showed that starting from a certain total angular momentum of the core, the angular momenta carried by the quadrupole ($\vec{J}_2$) and octupole ($\vec{J}_3$)  bosons respectively, are perpendicular on each other. Naturally, we may ask ourselves whether there exists a strength for  the particle-core interaction such that the angular momentum of the odd particle becomes perpendicular to the plane ($\vec{J}_2, \vec{J}_3$). This would be a signature that the three component system exhibits a chiral symmetry. 

As a final conclusion, one may say that the present CSM extension to odd nuclei can describe quite well the excitation energies in the
parity partner bands with $K^{\pi}=\frac{1}{2}^{\pm}$.

{\bf Acknowledgment.} This paper was supported by the Romanian Ministry of Education and Research under the contracts PNII, No. ID-33/2007
and ID-1038/2009. 

\renewcommand{\theequation}{A.\arabic{equation}}
\setcounter{equation}{0}

\section{Appendix A}
 The diagonal matrix elements of the quadrupole-quadrupole ($q_2Q_2$) and octupole-octupole ($q_3Q_3$) particle-core interactions in the basis defined in Section III are:
\begin{eqnarray}
&&\langle \varphi^{(\pm)}_{IM;j_iK}|q_2Q_2|\varphi^{(\pm)}_{IM;j_iK}=-X_2C^{j_i\;J\;I}_{k-2\;2\;K}
C^{j_i\;J^{'}\;I}_{k-2\;2\;K}\hat{I}^2\hat{j_i}\hat{J}W(j_iI2J;J^{'}j_i)\nonumber\\
&\times&\left(N^{(\pm)}_{I,K}\right)^2\left(N^{(\gamma,\pm)}_{J}N^{(\gamma,\pm)}_{J^{'}}\right)^{-1}\langle j_i||r^2Y_2||j_i\rangle\langle\varphi^{(\gamma;\pm)}_J||b^{\dagger}_2+b_2||
\varphi^{(\gamma;\pm)}_{J^{'}}\rangle,i=2,3; K=i-1/2,\nonumber\\
&&\langle \varphi^{(\pm)}_{IM;j_35/2}|q_3Q_3|\varphi^{(\pm)}_{IM;j_2 3/2}=X_3C^{j_3\;J\;I}_{1/2\;2\;5/2}C^{j_2\;J^{'}\;I}_{-1/2\;2\;3/2}\hat{I}^2\hat{j_3}\hat{J}W(j_2I3J;J^{'}j_3)\nonumber\\
&\times&N^{(\pm)}_{I,5/2}N^{(\pm)}_{I,3/2}\left(N^{(\gamma,\pm)}_{J}N^{(\gamma,\pm)}_{J^{'}}\right)^{-1}\langle j_3||r^3Y_3||j_i\rangle\langle\varphi^{(\gamma;\pm)}_J||b^{\dagger}_3+b_3||
\varphi^{(\gamma;\mp)}_{J^{'}}\rangle.
\end{eqnarray}
The expectation value for the $q_2Q_2$ term, in the state  $\varphi^{\pm}_{IM;j_11/2}$ can be obtained from the  expression  given above by replacing $j_i$ by $j_1$ and $\varphi^{(\gamma;\pm)}_J$ by $\varphi^{(g;\pm)}_J$. Also, the projections associated to $J$ and $J^{'}$ in the two Clebsch Gordan coefficients should be equal to zero and not 2.
It is easy to check that this state is not connected by the $q_3Q_3$ interaction to the state
$\varphi^{\pm}_{IM;j_35/2}$.
The reduced matrix elements of the boson operators involved in the above equations have the expressions:
\begin{eqnarray}
\langle \varphi^{(\gamma;\pm)}_J||b^{\dagger}_2+b_2||\varphi^{(\gamma;\pm)}\rangle &=&
dC^{J^{'}\; 2\; J}_{2\;0\;2}\left[\frac{N^{(\gamma;\pm)}_J}{N^{(\gamma;\pm)}_{J^{'}}}+
\frac{2J^{'}+1}{2J+1}\frac{N^{(\gamma;\pm)}_{J^{'}}}{N^{(\gamma;\pm)}_{J}}
+\frac{6}{7}\sum_{J^{'}}\frac{N^{(\gamma;\pm)}_{J}  N^{(\gamma;\pm)}_{J^{'}}}{N^{(g;\pm)}_{J_1}}
\right.\nonumber\\
&&\times\left.
\left((C^{J_1\;2\;J^{'}}_{0\;2\;2})^2+\frac{2J^{'}+1}{2J+1}(C^{J_1\;2\;J}_{0\;2\;2})^2\right)\right],
\nonumber\\
\langle \varphi^{(g;\pm)}_J||b^{\dagger}_2+b_2||\varphi^{(g;\pm)_{J^{'}}}\rangle &=&
dC^{J^{'}\; 2\; J}_{0\;0\;0}\left[\frac{N^{(g;\pm)}_J}{N^{(g;\pm)}_{J^{'}}}+
\frac{2J^{'}+1}{2J+1}\frac{N^{(g;\pm)}_{J^{'}}}{N^{(g;\pm)}_{J}}\right],\nonumber\\
\langle \varphi^{(\gamma;+)}_J||b^{\dagger}_3+b_3||\varphi^{(\gamma;-)}_{J^{'}}\rangle &=&
fC^{J^{'}\; 2\; J}_{2\;0\;2}\left[\frac{N^{(\gamma;+)}_J}{N^{(\gamma;-)}_{J^{'}}}+
\frac{2J^{'}+1}{2J+1}\frac{N^{(\gamma;-)}_{J^{'}}}{N^{(\gamma;+)}_{J}}\right],\nonumber\\
\langle \varphi^{(\gamma;-)}_J||b^{\dagger}_3+b_3||\varphi^{(\gamma;+)}_{J^{'}}\rangle &=&
(-)^{J^{'}-J}\frac{\hat{J^{'}}}{\hat{J}}\langle \varphi^{(\gamma;+)}_J||b^{\dagger}_3+b_3||\varphi^{(\gamma;-)}\rangle ,
\nonumber\\
\langle \varphi^{(g;+)}_J||b^{\dagger}_3+b_3||\varphi^{(g;-)}_{J^{'}}\rangle &=&
fC^{J^{'}\; 2\; J}_{0\;0\;0}\left[\frac{N^{(g;+)}_J}{N^{(g;-)}_{J^{'}}}+
\frac{2J^{'}+1}{2J+1}\frac{N^{(g;-)}_{J^{'}}}{N^{(g;+)}_{J}}\right],\nonumber\\
\langle \varphi^{(g;-)}_J||b^{\dagger}_3+b_3||\varphi^{(g;+)}_{J^{'}}\rangle &=&
(-)^{J^{'}-J}\frac{\hat{J^{'}}}{\hat{J}}\langle \varphi^{(g;+)}_{J^{'}}||b^{\dagger}_3+b_3||\varphi^{(g;-)}_{J}\rangle .
\end{eqnarray}
The matrix elements of $H_{core}$ have the expressions:
\begin{eqnarray}
\langle\varphi^{(\pm)}_{IM,j_23/2}|H_{core}|\varphi^{(\pm)}_{IM,j_23/2}\rangle &=&
N^{(\pm)}_{I;j_23/2}\sum_{J}\left(C^{j_2\;J\;I}_{-1/2\;2\;3/2}\right)^2\left(N^{(\gamma;\pm)}_J\right)^{-2}E^{(\gamma,\pm)}_{J},\nonumber\\
\langle\varphi^{(\pm)}_{IM,j_35/2}|H_{core}|\varphi^{(\pm)}_{IM,j_35/2}\rangle &=&
N^{(\pm)}_{I;j_35/2}\sum_{J}\left(C^{j_2\;J\;I}_{1/2\;2\;5/2}\right)^2\left(N^{(\gamma;\pm)}_J\right)^{-2}E^{(\gamma,\pm)}_{J},\nonumber\\
\langle\varphi^{(\pm)}_{IM,j_11/2}|H_{core}|\varphi^{(\pm)}_{IM,j_11/2}\rangle &=&
N^{(\pm)}_{I;j_11/2}\sum_{J}\left(C^{j_2\;J\;I}_{1/2\;0\;1/2}\right)^2\left(N^{(g;\pm)}_J\right)^{-2}E^{(g,\pm)}_{J},
\end{eqnarray}
where $E^{(g,\pm)}_J$ and $E^{(\gamma,\pm)}_J$ denote the energies of the state $J^{\pm}$ belonging
to the bands $g^{\pm}$ and $\gamma^{\pm}$, respectively.
Obviously, the term $H_{sp}$ is diagonal in the chosen basis:
\begin{eqnarray}
\langle\varphi^{(\pm)}_{IM,j_11/2}|H_{sp}|\varphi^{(\pm)}_{IM,j_11/2}\rangle &=&\epsilon_{j_1},
\nonumber\\
\langle\varphi^{(\pm)}_{IM,j_23/2}|H_{sp}|\varphi^{(\pm)}_{IM,j_23/2}\rangle &=&\epsilon_{j_2},
\nonumber\\
\langle\varphi^{(\pm)}_{IM,j_35/2}|H_{sp}|\varphi^{(\pm)}_{IM,j_35/2}\rangle &=&\epsilon_{j_3}.
\end{eqnarray}
Here $\epsilon_{j_k}$ denotes the energies of the spherical shell model states $|n_k,l_k,j_k,m_k\rangle$ with $k=1,2,3.$
The matrix elements of the term $\vec{j}\cdot\vec{J}$ are:
\begin{eqnarray}
\langle\varphi^{(\pm)}_{IM,j_11/2}|\vec{j}\cdot\vec{J}|\varphi^{(\pm)}_{IM,j_11/2}\rangle &=&
\frac{1}{2}\left[I(I+1)-j_1(j_1+1)-N^{(\pm)}_{I;j_11/2}\sum_{J}\left(C^{j_2\;J\;I}_{1/2\;0\;1/2}\right)^2\left(N^{(g;\pm)}_J\right)^{-2}J(J+1)\right],
\nonumber\\
\langle\varphi^{(\pm)}_{IM,j_23/2}|\vec{j}\cdot\vec{J}|\varphi^{(\pm)}_{IM,j_23/2}\rangle &=&
\frac{1}{2}\left[I(I+1)-j_2(j_2+1)-N^{(\pm)}_{I;j_23/2}\sum_{J}\left(C^{j_2\;J\;I}_{-1/2\;2\;3/2}\right)^2\left(N^{(\gamma;\pm)}_J\right)^{-2}J(J+1)\right],\nonumber\\
\nonumber\\
\langle\varphi^{(\pm)}_{IM,j_35/2}|\vec{j}\cdot\vec{J}|\varphi^{(\pm)}_{IM,j_35/2}\rangle &=&
\frac{1}{2}\left[I(I+1)-j_3(j_3+1)-N^{(\pm)}_{I;j_35/2}\sum_{J}\left(C^{j_2\;J\;I}_{1/2\;2\;5/2}\right)^2\left(N^{(\gamma;\pm)}_J\right)^{-2}J(J+1)\right].\nonumber\\
\end{eqnarray}

\renewcommand{\theequation}{B.\arabic{equation}}
\setcounter{equation}{0}

\section{Appendix B}
The matrix elements involved in the expression of the branching ratios are:
\begin{eqnarray}
\langle \varphi^{(\pi)}_I(j_iK;d,f)||r^2Y_{2}||\varphi^{(\pi)}_{I^{'}}(j_iK^{'};d,f)\rangle&=&
-\sqrt{\frac{5}{4\pi}}\langle r^2\rangle \hat{I^{'}}\hat{j_{i}} N^{(\pi)}_{i,IK}N^{(\pi)}_{i,I^{'}K^{'}}\\
&&\times\sum_{J}C^{j_i\;J\;I}
_{K\;0\;K}C^{j_i\;J\;I^{'}}_{K^{'}\;0\;K^{'}}\left(N^{(g,\sigma)}_J\right)^{-2},
\nonumber\\
\langle \varphi^{(\pi)}_I(j_iK;d,f)||b^{\dagger}_2+b_2||\varphi^{(\pi)}_{I^{'}}(j_iK^{'};d,f)\rangle&=&
dC^{I^{'}\;2\;I}_{K\;0\;K}\left(\frac{N^{(\pi)}_{i;IK}}{N^{(\pi)}_{i;I^{'}K^{'}}}
+\frac{2I^{'}+1}{2I+1}\frac{ N^{(\pi)}_{i;I^{'}K^{'}}}{ N^{(\pi)}_{i;IK}}\right),
\nonumber\\
\langle \varphi^{(\pi)}_I(j_iK;d,f)||\left(b^{\dagger}_2b^{\dagger}_3\right)_1+\left(b_3b_2\right)_1||\varphi^{(\pi^{'})}_{I^{'}}(j_iK^{'};d,f)\rangle&=&
dfC^{I^{'}\;1\;I}_{K\;0\;K}C^{2\;3\;1}_{0\;0\;0}\left[\frac{ N^{(\pi)}_{i;IK}}{ N^{(\pi^{'})}_{i;I^{'}K^{'}}}+\frac{2I^{'}+1}{2I+1}\frac{ N^{(\pi^{'})}_{i;I^{'}K^{'}}}{ N^{(\pi)}_{i;IK}}\right]
\nonumber.
\end{eqnarray}

\end{document}